# Level densities and thermodynamical quantities of heated $^{93-98}$Mo isotopes

R. Chankova *,[1] A. Schiller,[2] U. Agvaanluvsan,[2, 3] E. Algin,[2, 3, 4, 5] L.A. Bernstein,[2] M. Guttormsen,[1] F. Ingebretsen,[1] T. Lönnroth,[6] S. Messelt,[1] G.E. Mitchell,[3, 4] J. Rekstad,[1] S. Siem,[1] A.C. Larsen,[1] A. Voinov,[7, 8] and S. Ødegård[1]

[1]*Department of Physics, University of Oslo, N-0316 Oslo, Norway*
[2]*Lawrence Livermore National Laboratory, L-414, 7000 East Avenue, Livermore, CA 94551, USA*
[3]*North Carolina State University, Raleigh, NC 27695, USA*
[4]*Triangle Universities Nuclear Laboratory, Durham, NC 27708, USA*
[5]*Department of Physics, Osmangazi University, Meselik, Eskisehir, 26480 Turkey*
[6]*Department of Physics, Åbo Akademi, FIN-20500 Turku, Finland*
[7]*Department of Physics and Astronomy, Ohio University, Athens, Ohio 45701, USA*
[8]*Frank Laboratory of Neutron Physics, Joint Institute of Nuclear Research, 141980 Dubna, Moscow region, Russia*

Level densities for $^{93-98}$Mo have been extracted using the ($^3$He,$\alpha\gamma$) and ($^3$He,$^3$He'$\gamma$) reactions. From the level densities thermodynamical quantities such as temperature and heat capacity can be deduced. Data have been analyzed by utilizing both the microcanonical and the canonical ensemble. Structures in the microcanonical temperature are consistent with the breaking of nucleon Cooper pairs. The S-shape of the heat capacity curves found within the canonical ensemble is interpreted as consistent with a pairing phase transition with a critical temperature for the quenching of pairing correlations at $T_c \sim 0.7 - 1.0$ MeV.

## I. INTRODUCTION

Level density is a characteristic property of many-body quantum mechanical systems. Its precise knowledge is often a key ingredient in the calculation of different processes, such as compound nuclear decay rates, yields of evaporation residues to populate exotic nuclei, or thermonuclear rates in astrophysical processes.

Measurements of experimental nuclear level density are an important prerequisite for thermodynamical studies of atomic nuclei. Level density is directly connected to the multiplicity of states, i.e., the number of physical realizations of the system at a certain excitation energy. The entropy is a fundamental quantity and a measure of the disorder of the many-body system. Within the microcanonical ensemble it is defined as the natural logarithm of the multiplicity of states. When the entropy is known, thermodynamical quantities such as temperature and heat capacity can be extracted. These quantities depend on the statistical properties of the nuclear many-body system and may reveal phase transitions.

Pairing correlations are one of the fundamental properties of nuclei, and have been successfully described by the Bardeen-Cooper-Schrieffer (BCS) theory of superconductivity [1]. By using the BCS theory the thermodynamical properties of nuclear pairing were investigated in the study of warm nuclei [2, 3, 4, 5]. In the case of a finite Fermi system such as the nucleus, statistical fluctuations beyond the mean field become important. The fluctuations smooth out the sharp phase transition, and then the pairing correlations do not vanish suddenly but decrease with increasing temperature. The quenching of pair-ing correlations has been obtained in recent theoretical approaches: the shell model Monte Carlo (SMMC) calculations [6, 7, 8], the finite-temperature Hartree-Fock-Bogoliubov theory [9], and the relativistic mean field theory [10]. Experimental data on the quenching of pair correlations are important as a test for nuclear theories. A long-standing problem in experimental nuclear physics has been to observe the transition from strongly paired states at zero temperature to unpaired states at higher temperatures. A signature of the pairing transition at finite temperature might be a local increase in the heat capacity as a function of temperature [11]. Recently [12, 13], fine structures in the level densities in the 1–7 MeV region were reported, which are probably due to the breaking of individual nucleon pairs and a gradual decrease of pair correlations.

The group at the Oslo Cyclotron Laboratory (OCL) has developed a method to extract simultaneously the level density and the radiative strength function from primary $\gamma$ spectra [14]. The method is a further development of the sequential extraction method [15, 16]. The Oslo method has been tested in the rare-earth mass region which led to many interesting applications [12, 17, 18, 19]. In order to make quantitative judgments of the applicability of the method, the Oslo Cyclotron group has extracted the level density and radiative strength function (RSF) of the very light $^{27,28}$Si nuclei, where these quantities are known. Excellent overall agreement was found [20]. Subsequently, another extension has been made to the intermediate nuclei $^{56,57}$Fe and $^{96,97}$Mo, and it has been shown that the method can be applied in this intermediate mass region where the level density is still relatively low [21, 22]. All of these successful applications have motivated us to employ the Oslo method to study medium-heavy nuclei in the vicinity of closed shells.

*Electronic address: rositsa.chankova@fys.uio.no

The naturally occurring isotopes of molybdenum span one of the larger isotopic ranges and are well suited as targets for the study of nuclear properties, such as the effect of changing from spherical to deformed shapes. When approaching closed shells, the nuclear structure changes significantly, and one expects this to influence the level densities and radiative strength functions.

The even-even $^{92}$Mo has a filled $N = 50$ neutron shell [23]. It is essentially a spherical nucleus and vibrations are primarily governed by the proton core. As the mass increases from $^{94}$Mo to $^{100}$Mo, neutrons fill the $2d_{5/2}$ and $1g_{7/2}$ sub-shells. Moving away from the $N = 50$ shell closure, pairing and quadrupole interactions cause a more collective behavior in the heavier Mo isotopes. The character of the isotopes changes rapidly from that of the essentially spherical $^{92}$Mo to nuclei making a transition from collective vibrators to the deformed rotors of the unstable $^{104}$Mo and $^{106}$Mo isotopes [24]. The transitional nature of molybdenum isotopes away from $N = 50$ has been the focus of several efforts as described in [25] and references therein.

Around closed shells, effects from the increasing single particle energy spacings can be expected. These will also influence the entropy difference between odd-mass and even-even nuclei. Therefore, a statistical description of the transition from closed shells to deformed nuclei is of great interest.

In this work, a unique and consistent investigation of the six $^{93-98}$Mo isotopes is performed in order to determine experimentally the level density from the ground state up to the neutron binding energy. The Oslo method also determines the RSFs of the molybdenum isotopes studied; these are presented in an earlier paper [26].

## II. EXPERIMENTAL METHODS

The experiments were carried out at the Oslo Cyclotron Laboratory by bombarding $^{94,96,97,98}$Mo targets with $^{3}$He ions. In the present work, results from eight different reactions on four different targets are discussed. These are the following six reactions which are the subject of the present investigation:
1. $^{98}$Mo($^{3}$He,$\alpha\gamma$)$^{97}$Mo (45 MeV)
2. $^{98}$Mo($^{3}$He,$^{3}$He$'\gamma$)$^{98}$Mo (45 MeV)
3. $^{96}$Mo($^{3}$He,$\alpha\gamma$)$^{95}$Mo (30 MeV)
4. $^{96}$Mo($^{3}$He,$^{3}$He$'\gamma$)$^{96}$Mo (30 MeV)
5. $^{94}$Mo($^{3}$He,$\alpha\gamma$)$^{93}$Mo (30 MeV)
6. $^{94}$Mo($^{3}$He,$^{3}$He$'\gamma$)$^{94}$Mo (30 MeV)
together with the reactions
7. $^{97}$Mo($^{3}$He,$\alpha\gamma$)$^{96}$Mo (45 MeV)
8. $^{97}$Mo($^{3}$He,$^{3}$He$'\gamma$)$^{97}$Mo (45 MeV)
which have been reported earlier [21, 22]. The self-supporting targets with thicknesses of $\sim 2$ mg/cm$^2$ are enriched to $\sim 95\%$. The experiments were run with beam currents of $\sim 2$ nA for 1–2 weeks. The particle-$\gamma$ coincidences were measured with the CACTUS multi-detector array. The charged ejectiles were detected by eight par-ticle telescopes placed at an angle of $45°$ relative to the beam direction. An array of 28 NaI $\gamma$-ray detectors with a total efficiency of $\sim 15\%$ surrounded the target and particle detectors.

For each initial excitation energy, the $\gamma$-ray spectra are recorded as a function of the initial excitation energy of the residual nucleus. This is accomplished by utilizing the known reaction Q-values and kinematics. Using the particle-$\gamma$ coincidence technique, each $\gamma$ ray can be assigned to a cascade depopulating a certain initial excitation energy in the residual nucleus. The data are therefore sorted into total $\gamma$-ray spectra originating from different initial excitation-energy bins. Each spectrum is then unfolded with the NaI response function using a Compton-subtraction method which preserves the fluctuations in the original spectra and does not introduce further, spurious fluctuations [27]. From the unfolded spectra, a primary-$\gamma$ matrix $P(E, E_\gamma)$ is constructed using the subtraction method of Ref. [28].

The basic assumption of this method is that the $\gamma$-ray energy distribution from any excitation energy bin is independent of whether states in this bin are populated directly via the ($^{3}$He,$\alpha$) or ($^{3}$He,$^{3}$He$'$) reactions or indirectly via $\gamma$ decay from higher excited levels following the initial nuclear reaction. This assumption is trivially fulfilled if one populates the same levels with the same weights within any excitation energy bin, since the decay branchings are properties of the levels and do not depend on the population mechanisms. The assumptions behind this method have been tested extensively by the Oslo group and have been shown to work reasonably well [29].

The ($^{3}$He,$^{3}$He$'\gamma$) and ($^{3}$He,$\alpha\gamma$) reactions exhibit very different reaction mechanisms. This is demonstrated in Fig. 1 where the particle spectra in coincidence with $\gamma$ rays show indeed very different yields and peak structures.

The ($^{3}$He,$\alpha\gamma$) pick-up reaction reveals a cross section dominated by high $\ell$ neutron transfer. Here, the direct population of the residual nucleus takes place through one-particle-one-hole components of the wave function. Such configurations are not eigenstates of the nucleus, but they are rather distributed over virtually all eigenstates in the neighboring excitation-energy region. Thus, the neutron-hole strength for single particle levels away from the Fermi energy is distributed over a rather large range of background states.

On the other hand the inelastic scattering ($^{3}$He,$^{3}$He$'\gamma$) reaction is known to populate mainly collective excitations with a slightly lower spin window. Collective excitations built on the ground state give rise to rather pure eigenfunctions and their strength is less spread over other eigenfunctions of the nucleus in the neighboring excitation energy region.

In order to test if the number of $\gamma$-rays per cascade depends on the two types of reactions, we have evaluated the average $\gamma$-ray multiplicity

$$\langle M_\gamma(E)\rangle = \frac{E}{\langle E_\gamma\rangle}. \quad (1)$$

as a function of excitation energy $E$. The average $\gamma$-ray energy $\langle E_\gamma\rangle$ is calculated from $\gamma$ spectra selected at a certain energy $E$.

Figure 2 shows the $\gamma$-ray multiplicity versus excitation energy. In spite of the different reaction mechanisms, the two reactions give similar results. In particular, the multiplicities (solid and dashed lines) of $^{96}$Mo and $^{97}$Mo are equal within their error bars, which gives support to the applicability of the Oslo method for both reactions.

The experimental extraction procedure and assumptions of the Oslo method are given in Refs. [14, 29] and references therein. The first generation (or primary) $\gamma$-ray matrix which is obtained as described above can be factorized according to the Brink-Axel hypothesis [30, 31] as

$$P(E,E_\gamma) \propto \rho(E-E_\gamma)\mathcal{T}(E_\gamma), \quad (2)$$

where $\rho$ is the level density and $\mathcal{T}$ is the radiative transmission coefficient.

The $\rho$ and $\mathcal{T}$ functions can be determined by an iterative procedure [14] through the adjustment of each data points of these two functions until a global $\chi^2$ minimum with the experimental $P(E,E_\gamma)$ matrix is reached. It has been shown [14] that if one solution for the multiplicative functions $\rho$ and $\mathcal{T}$ is known, one may construct an infinite number of other functions, which give identical fits to the $P$ matrix by

$$\tilde{\rho}(E-E_\gamma) = A\exp[\alpha(E-E_\gamma)]\,\rho(E-E_\gamma), \quad (3)$$

$$\tilde{\mathcal{T}}(E_\gamma) = B\exp(\alpha E_\gamma)\mathcal{T}(E_\gamma). \quad (4)$$

Consequently, neither the slope nor the absolute values of the two functions can be obtained through the fitting procedure. Thus, the parameters $\alpha$, $A$ and $B$ remain to be determined.

The parameters $A$ and $\alpha$ can be determined by normalizing the level density to the number of known discrete levels at low excitation energy [32] and to the level density estimated from neutron-resonance spacing data at the neutron binding energy $E = B_n$ [33]. The procedure for extracting the total level density $\rho$ from the resonance energy spacing $D$ is described in Ref. [14]. Since our experimental level-density data points reach up to an excitation energy of only $E \sim B_n - 1$ MeV, we extrapolate with the back-shifted Fermi-gas model [34, 35]

$$\rho_{\mathrm{BSFG}}(E) = \eta\frac{\exp(2\sqrt{aU})}{12\sqrt{2}a^{1/4}U^{5/4}\sigma_I}, \quad (5)$$

where a constant $\eta$ is introduced to adjust $\rho_{\mathrm{BSFG}}$ to the experimental level density at $B_n$. The intrinsic excitation energy is estimated by $U = E - C_1 - E_{\mathrm{pair}}$, where $C_1 = -6.6A^{-0.32}$ MeV and $A$ are the back-shift parameter and mass number, respectively. The pairing energy

$E_{\mathrm{pair}}$ is based on pairing gap parameters $\Delta_p$ and $\Delta_n$ evaluated from even-odd mass differences [36] following the prescription of Dobaczewski $et\ al.$ [37]. The level density parameter is given by $a = 0.21A^{0.87}$ MeV$^{-1}$. The spin-cutoff parameter $\sigma_I$ is given by $\sigma_I^2 = 0.0888aTA^{2/3}$, where the nuclear temperature is given by

$$T = \sqrt{U/a}. \quad (6)$$

In cases where the intrinsic excitation energy $U$ becomes negative, we set $U = 0$, $T = 0$ and $\sigma_I = 1$.

Figure 3 demonstrates the level density normalization procedure for the $^{97}$Mo case. The experimental data points are normalized according to (3) by adjusting the parameters $A$ and $\alpha$ such that a least $\chi^2$ fit is obtained in between the arrows. For the lower excitation region (see upper panel), one should take care only to include a fit region where it is likely that (almost) all levels are known. In practice, this means that the level density should not exceed $\sim 50$ levels per MeV. At the higher excitation region (lower panel), the Fermi-gas extrapolation connects seamlessly the highest-energy data points with the level-density value determined from neutron-resonance spacing at $B_n$. Generally, the resulting normalization is not very dependent on the choice of the theoretical extrapolation function (Fermi gas) or the chosen fit region ($\sim 4.5$ to $\sim 7$ MeV).

Unfortunately, no information exists on the level density at $E = B_n$ for $^{94}$Mo. Therefore, we estimate this value from a systematics of other Mo isotopes where information on the level density at $B_n$ exists. In Fig. 4 we plot all the known data points from the Mo isotopic chain. The odd- and even-mass molybdenum nuclei fall into two groups, both showing a decreasing level density as function of excitation energy. This behavior is rather counterintuitive since in a given nucleus the level density increases exponentially with excitation energy, and for neighboring nuclei one would naively expect quite similar level-density curves. Two effects combine to result in the negative slope of the data points: (i) the decrease of single-particle level density when approaching the $N = 50$ shell gap resulting in a decrease of the level density in general, and (ii) the increase of the neutron-separation energy with decreasing neutron number. For the negative slope to emerge, both effects have to be rather precisely of the same size for each step along the Mo isotopic chain. We have found no good physical explanation for this to happen, but we employ this fortuitous coincidence to estimate $\rho(B_n) = (6.2\pm1.0)10^4$ MeV$^{-1}$ for $^{94}$Mo from our phenomenological systematics[1]. The splitting of data points between even and odd Mo isotopes must not be interpreted solely as due to the effect of the pairing-energy shift of the level-density curves; the difference in the mag-nitude of $B_n$ between neutron-odd and -even isotopes also affects the magnitude of this splitting.

---

[1] This value also agrees within a factor of two with the systematics of Ref. [38]

## III. LEVEL DENSITY AND FINE STRUCTURES OF THE ENTROPY

The present knowledge on level density is concentrated in mainly two regions; the low excitation region up to $\sim 2$ MeV, studied in detail using spectroscopy and counting of known, discrete levels [39] and the region around the neutron separation energy studied by experiments on neutron resonances [40]. Almost nothing is known of the level density in between the above mentioned regions, but it is possible to determine quite reliably two anchor points of the level density.

Figure 5 shows the extracted anchor points (filled data points) for nine molybdenum isotopes together with the level density deduced from known discrete levels (solid lines). The upper anchor point is simply determined from neutron resonance data. The lower anchor point, which is the average value of three data points, is determined such that a straight line on a logarithmic plot, going through the upper anchor point, provides an upper bound of the level density distribution of known levels. The algorithm is iterative and treats all nuclei similarly to ensure that the results are comparable. The straight line connecting the lower and upper anchor points, is defined by the constant temperature formula

$$\rho(E) = Ce^{E/\tau} \tag{7}$$

with $\tau^{-1} = (\ln \rho_2 - \ln \rho_1)/(E_2 - E_1)$ and $C = \rho_1 \exp(-E_1/\tau)$. Details are given in [41]. Provided that all the levels are measured at the excitation energy of the lower anchor point, we find from the plots of Fig. 5 that the temperature-like parameter $\tau$ drops from 1.05 MeV for the spherical $^{93}$Mo to about 0.72 MeV for the well deformed $^{101}$Mo nucleus. This figure also illustrates the excitation energy where one would expect the appearance of missing levels in spectroscopic work, typically if the density of levels exceeds 50 MeV$^{-1}$.

The level densities $\rho(E)$ extracted from the eight reactions are displayed in Fig. 6. The data have been normalized as prescribed above, and the parameters used for $^{93-98}$Mo in Eq. (5) are listed in Table I. We find that the normalization constant $\eta$ drops by one order of magnitude when going from deformed to spherical nuclei. This means that the spherical $^{93}$Mo has about ten times lower level density than predicted by a global Fermi-gas model. As mentioned earlier, this effect is one of the reasons for the negative slope of the data points in Fig. 4.

Our experimental data interpolate between the previously known lower anchor point at $\sim 2$ MeV and about 1 MeV below the upper anchor point at $\sim 7$ MeV. For the energy interval between $\sim 6$ MeV and $\sim 7$ MeV, we rely on models [34, 35]. In spite of the fact, that the final extrapolation of the level density up to the nuclear binding energy is model dependent, this only affects the average slope of the level density and does not affect the fine structure. This enables us to observe fine structures in the level density which are thought to reflect the breaking of individual pairs. In an earlier work, we showed how a simple single-particle plus pairing model can qualitatively explain the emergence of such fine structures [21]. Moreover, we have in the past investigated how pairing correlations are weakened in the presence of already unpaired nucleons, but also how these unpaired nucleons around the Fermi energy can increase the cost in energy to break up further nucleon pairs due to the blocking effect of the Pauli principle [42]. Our goal in the present work is to obtain experimental values for the critical temperature of the pair-breaking process. On the way, we will also investigate some other thermodynamical properties, in particular the entropy, when going from spherical to deformed nuclei. The generalization of the concept of temperature for a small system is not straightforward and has been discussed extensively in the literature (see, e.g., Ref. [42] and references therein). Traditionally, temperature is introduced in slightly different ways in the microcanonical statistical ensemble (as a property of the system itself) and in the canonical statistical ensemble (as imposed by a heat bath). The temperature-energy relations for rare-earth nuclei (the caloric curves) derived within the two statistical ensembles display in general a different behavior since the nuclei under discussion are essentially discrete systems [13].

In order to avoid the shortcomings imposed by the above mentioned statistical ensembles, a new approach for the caloric curves based on the two-dimensional probability distribution $P(E,T)$ has been proposed [42, 43]. This approach bypasses the well-known problem of spurious structures such as negative temperatures and heat capacities in the microcanonical ensemble. On the other hand more structures in the new caloric curve are evident than in the canonical caloric curve. However, this new method is still not well settled and we will defer the discussion of caloric curves to another occasion.

Within the microcanonical ensemble the experimentally obtained level density corresponds to the partition function which is simply the multiplicity $\Omega$ of accessible states. Thus, the entropy $S$ of the system within the energy interval $E$ and $E + \delta$ is determined by:

$$S(E) = k_{\mathrm{B}} \ln \Omega(E) \tag{8}$$

where $\Omega(E) = \rho(E)/\rho_0$ and the Boltzmann constant is set to unity ($k_B \equiv 1$) for simplicity[2]. In order to fulfill the third law of thermodynamics; namely $S \to 0$ when $T \to 0$, the normalization denominator is set to $\rho_0 = 1.5$ MeV$^{-1}$. Entropy as a function of energy can be defined and measured for small and mesoscopic systems as well as for large systems. However, fluctuations in level spacings which are typical for small systems will make the entropy sensitive to exactly how the energy interval between $E$ and $E + \delta E$ is chosen. Thus, Eq. (8) is only useful if $\Omega(E)$ is a sufficiently smooth function, i.e., for the case that its first derivative exists. Small statistical fluctuations in the entropy $S$ may give rise to large contributions to the temperature $T$, which is defined within the microcanonical ensemble as

$$T(E) = \left( \frac{\partial S}{\partial E} \right)^{-1}. \qquad (9)$$

---

[2] More precisely, multiplicity $\Omega(E)$ is proportional to $\rho(E)(2\langle J(E)\rangle + 1)$, where $\langle J(E)\rangle$ is the average spin of levels with excitation energy $E$. However, in the present work, we neglect the weak excitation-energy dependence of $\langle J(E)\rangle$.

Figure 7 shows the entropy deduced within the microcanonical ensemble for $^{93,94}$Mo (upper panel) and $^{97,98}$Mo (lower panel). The entropy curve plotted on a linear scale is essentially identical to the level-density curve on a logarithmic scale. In general, the most efficient way to create additional states in atomic nuclei is to break $J = 0$ nucleon Cooper pairs from the core. The resulting two nucleons may thereby be thermally excited rather independently to the available single particle levels around the Fermi surface. We therefore interpret, e.g., the steplike increases in entropy around 2–3 MeV excitation energy in Fig. 7 as due to the breaking of the first nucleon Cooper pair.

The entropies of odd-mass nuclei are higher than those of their even-even neighbors, even when their mass numbers are lower. It is interesting to compare entropies between neighboring nuclei. The difference in entropy

$$\Delta S(E) = S_{\text{odd-mass}} - S_{\text{even-even}} \qquad (10)$$

is assumed to be a measure for the single particle entropy. The entropies of the almost spherical $^{93}$Mo and $^{94}$Mo (lower panel of Fig. 7) follow each other closely above $E \sim 2.5$ MeV. Here, the odd valence nucleon behaves almost as a passive spectator. For $^{93,94}$Mo, we find $\Delta S \gtrsim 0$ for $E > 2.5$ MeV. The deformed case, (lower panel of Fig. 7) exhibits an entropy difference of $\Delta S \gtrsim 1$. This is less than the value of $\Delta S \sim 2$ found for rare-earth nuclei [44, 45].

These observations can be explained qualitatively by the fact that the single particle entropy depends on the number of single particle orbitals that are available for excitations at a certain temperature. For $^{93,94}$Mo at low energies, the single neutron outside the closed shell can only occupy the two $d_{5/2}$ and $g_{7/2}$ orbitals giving an entropy of $\ln 2 \sim 0.7$. For the deformed nucleus $^{97,98}$Mo, symmetry breaking results in a splitting of these two single particle orbitals into seven Nilsson orbitals, giving a total entropy of $\ln 7 \sim 1.9$, i.e., about one unit more than for the $^{93,94}$Mo case. In the rare-earth region strong deformation and intruder orbitals from other shells result in an even higher single particle level density, giving rise to an even larger single particle entropy. Although our simple argument somewhat overestimates the observed single particle entropies, it provides a satisfactory explanation for the differences between the single particle entropies in the different cases.

The entropy in atomic nuclei at low energies does not simply scale with the total number of nucleons. In the presence of pairing correlations, i.e., away from closed shells, the entropy scales rather with the number of unpaired nucleons at a certain excitation energy. When pairing correlations cannot form due to the large single particle level spacings around closed shells, an unpaired nucleon will behave almost as a passive spectator without contributing significantly to the entropy of the system.

Around 5.5 MeV excitation energy, a bump is observed in the entropy curves for the lighter $^{93,94}$Mo nuclei. In light of the discussion above, it is unlikely that such a bump can be interpreted as the breaking of a nucleon Cooper pair. We propose that this bump is related to the sudden onset of neutron excitations across the $N = 50$ shell gap. Due to the generally higher level density and the onset of deformation in the heavier Mo isotopes, the opening of the $g_{9/2}$ shell might be less significant, leading to the effect being spread out in energy. However, such an effect might become visible in the lighter $^{93,94}$Mo nuclei. This interpretation is supported by the fact that the large transfer peak at 5.5 MeV excitation energy in the particle spectrum of the $^{97}$Mo($^{3}$He,$\alpha\gamma$)$^{96}$Mo reaction at 45 MeV beam energy (see Fig. 1) has been shown in an experiment at the Yale University Enge splitpole to be dominated by high $\ell$ transfer, most likely $\ell = 4\hbar$ [46].

## IV. PHASE TRANSITIONS

### A. Model

In this Section we will utilize a recently developed thermodynamic model [41, 47, 48] which allows the investigation and classification of the pairing phase transition. The model is based on the canonical ensemble theory where equilibrium is obtained at a certain given temperature $T$. It can describe level densities for mid-shell nuclei in the mass regions $A = 58$, 106, 162, and 234.

The basic idea of the model is the assumption of a reservoir of nucleon pairs. These nucleon pairs can be broken and the unpaired nucleons are thermally scattered into an infinite, equidistant, doubly degenerated single-particle level scheme. The nucleon pairs in the reservoir do not interact with each other and are thought to occupy an infinitely degenerated ground state. The nucleons in the single particle level scheme do not interact with each other either, but they have to obey the Pauli principle.

The essential parameters of the model are: the number of pairs $n$ in the reservoir at zero temperature, the spacing of the single-particle level scheme $\epsilon = 30$ MeV/$A$, and the energy necessary to break a nucleon pair $2\Delta = 24$ MeV/$\sqrt{A}$. Quenching of pairing correlations is introduced in this model by reducing the energy required to break a nucleon pair in the presence of unpaired nucleons. We assume that for every already broken nucleon pair, the energy to break a further nucleon pair is reduced by a factor of $r = 0.56$ which is suggested by theoretical calculations [49]. In order to simulate the effect of the $N = 50$ shell closure on the $A < 98$ isotopes, we depart from the global systematics for $\epsilon$ and replace it by $\epsilon' = \epsilon a(A = 98)/a(A < 98)$ using the experimentally deduced $a$ values of Ref. [40]. We use the same parameters for both protons and neutrons, keeping the proton pairs fixed to seven because there are fourteen more protons outside the $Z = 28$ shell closure.

The total partition function is written as a product of proton ($Z_\pi$), neutron ($Z_\nu$), rotation ($Z_{\rm rot}$), and vibration ($Z_{\rm vib}$) partition functions where the parameters for the collective excitations are the rigid moment of inertia $A_{\rm rig} = 34$ MeV $A^{-5/3}$ and the energy of one-phonon vibrations $\hbar\omega_{\rm vib} = 1.5$ MeV taken from spectroscopic data [39]. Thermodynamical quantities of interest can be deduced from the Helmholtz free energy defined as

$$F(T) = -T\ln\left(Z_\pi Z_\nu Z_{\rm rot} Z_{\rm vib}\right). \tag{11}$$

This equation connects statistical mechanics and thermodynamics. Quantities such as entropy, average excitation energy, and heat capacity can be calculated by

$$S(T) = -\left(\frac{\partial F}{\partial T}\right)_V \tag{12}$$

$$\langle E(T)\rangle = F + ST \tag{13}$$

$$C_V(T) = \left(\frac{\partial\langle E\rangle}{\partial T}\right)_V, \tag{14}$$

respectively.

In Fig. 8, the Helmholtz free energy $F$, entropy $S$, average excitation energy $\langle E\rangle$, and heat capacity $C_V$ are shown as functions of temperature for even-even, odd and odd-odd systems in the $^{96}$Mo mass region. The free energy $F$ and the average excitation energy $\langle E\rangle$ are rather structureless as functions of temperature. The odd-even effects are small: The even-even, odd and odd-odd systems have different excitation energies at the same temperature, where the even-even system requires the highest $\langle E\rangle$ to be heated to some given temperature $T$. Around $T_c \sim 0.9$–$1.1$ MeV the nuclei are excited to their respective nucleon separation energies.

The entropy $S$ and heat capacity $C_V$ are more sensitive to thermal changes. The entropy difference $\Delta S$ between systems with $A$ and $A \pm 1$ is a useful quantity. At moderate temperatures, it is approximately extensive (additive) and represents the single-particle entropy associated with the valence particle (or hole) [41]. In the upper right panel, we find, e.g., that the nucleon carries a single-particle entropy of $\Delta S \sim 2.0$ at $T \sim 0.4$ MeV.

The shape of the heat-capacity curve is related to the level density. Traditionally, level-density curves have been described by the two component model as proposed by Gilbert and Cameron [34]. Within this model, the low energetic part is a constant-temperature level density and the high energetic part is a Fermi-gas model.

It has been shown in Ref. [12] that the inclusion of a constant-temperature part in the level-density formula creates a heat-capacity curve as function of temperature with a pronounced $S$ shape similar to that shown in Fig. 8. With our model parameters, the maximum of the local increase in the $C_V$ curve takes place at about $T \sim 0.9$ MeV. This temperature compares well with the temperature determined in the microcanonical ensemble from Eq. (9), giving a temperature of $T \sim 0.9$ MeV for $^{96}$Mo (see also Fig. 5).

### B. Comparison with experimental data

Our model is described within the canonical ensemble while experimental data refer to the microcanonical ensemble. There are two ways to compare our model with experiments. Details are given in [41]. In this work we will make use of the saddle-point approximation [50]

$$\rho(\langle E\rangle) = \frac{\exp(S)}{T\sqrt{2\pi C_V}}, \tag{15}$$

which gives satisfactory results for the nuclear level density [41, 48].

Figure 9 shows the theoretical level densities calculated using Eq. (15). The agreement with the anchor points and the experimental level densities for $^{97,98}$Mo isotopes is satisfactory. Some of the model parameters could be adjusted more precisely, however, in this work we have chosen to use parameters taken from systematics.

In order to investigate the behavior of the pairing correlations when approaching a major shell gap, we compare the canonical $C_V$ curves which are based on the Laplace transforms of the experimental level densities. The curves are plotted on Fig. 10 for even $^{94,96,98}$Mo (upper panel) and odd $^{93,95,97}$Mo (lower panel) nuclei. The $C_V$ curves resemble washed-out step structures and show an $S$ shape as a function of temperature quite similar to the model calculation on the lower right panel of Fig. 8. Because of the averaging performed by the Laplace transformation discrete transitions between the different quasi-particle regimes, as observed within the micro-canonical ensemble, are hidden. Only the phase transition related to the quenching of the pair correlations as a whole can be seen. Details are given in [17].

The canonical heat-capacity curves show local enhancements around $T \sim 0.5$–$1.0$ MeV. Such enhancements were predicted in the calculations of Fig. 8, and they are expected to be larger in the even-mass nuclei compared to the odd-mass neighbors. The experimental heat capacities show this feature for the $^{97,98}$Mo pair, and up to a certain extend for the $^{93,94}$Mo pair, but it is not very obvious for the $^{95,96}$Mo pair, where $^{95}$Mo shows a more pronounced enhancement than expected. Approaching the $N = 50$ closed shell, the local enhancements become less and less pronounced. The general behavior of pairing correlations is that at shell closure there are almost no pairing correlations and, as particles are added, the pairing correlations increase. Therefore the signature of a transition from a 'paired phase' to an 'unpaired phase' when approaching a major shell gap becomes less and less pronounced. We should note that very recently an alternative interpretation has been given [51]. These authors find that the S-shape can be accounted for as an effect of the particle-number conservation, and it occurs even when assuming a constant gap in the BCS theory.

From the $C_V$ curves, we have extracted the critical temperature for the quenching of pair correlations. The critical temperatures have been obtained by a fit of the canonical heat capacity of a constant-temperature level-density model to the data for the first 600 keV in temperature. The algorithm and its sensitivity are discussed in Ref. [12]. The values obtained are plotted in Fig. 11; there is a tendency for the critical temperature to be slightly higher for odd $^{93,95,97}$Mo than for even $^{92,94,96}$Mo nuclei, similar to the local enhancement of the heat-capacity curve in the model calculation (see the lower right panel of Fig. 8) which is observed at higher temperatures for odd-mass Mo isotopes. The higher critical temperature for odd-mass nuclei is due to the Pauli blocking effect of the unpaired quasiparticle which increases the distance to the Fermi surface for low-lying orbitals with coupled pairs and thus, increases the cost in energy to break pairs into more quasiparticles. Incidentally, the critical temperature for the quenching of pairing correlations coincides (by construction) quite well with the temperature-like parameter $\tau$ of Fig. 5.

A discontinuity of the heat capacity in a macroscopic system indicates a second-order phase transition according to the Ehrenfest classification; this is observed in the transition of a paired Fermion system such as a low-temperature superconductor or superfluid $^3$He to their normal phases. Thus, the experimentally observed local enhancement of the heat capacity is interpreted as a fingerprint of a phase transition from a phase with strong pairing correlations to a phase where the pairing correlations are quenched [12]. Shell-model Monte-Carlo calculations [7] have shown that the pairing-phase transition is strongly correlated with the suppression of neutron pairs with increasing temperature. It has also been observed that the reduction of the neutron-pair content of the wavefunction is much stronger in the even-even than in the odd-mass isotopes, giving rise to the more pronounced $S$ shape in the canonical heat-capacity curves in the even-even nuclei. The same odd-even difference in the heat capacity is also observed experimentally between $^{161}$Dy and $^{162}$Dy, and $^{171}$Yb and $^{172}$Yb [12].

## C. Entropy as function of neutron number

In order to study entropy as a function of neutron number, we compare the micro-canonical entropy obtained by the saddle-point approximation of Eq. (15) to our experimental data. In Fig. 12 the data are plotted as a function of the neutron number $N$ (left panel) and as a function of the number of available neutrons in the reservoir (right panel). Although only qualitative agreement is achieved, some simple conclusions can be drawn.

For the isotopes under investigation in this work, we see that the entropy at 1 MeV in both panels increases moderately as a function of the number of particles. The entropy at 7 MeV increases more rapidly and this is correlated to the evolution of the temperature-like parameter $\tau$, see Fig. 5. Both theoretically and experimentally, the odd systems show $\Delta S = 1.0k_B$ higher entropy than their neighboring even-even systems.

The slopes at 7 MeV in the left panel of Fig. 12 give $dS/dN = 0.5k_B$. Thus, going from $^{98}$Mo to $^{93}$Mo the level density drops when approaching the $N = 50$ shell gap by a factor of $\sim 0.03$. This mechanism is also reflected in the $\eta$ parameter of Eq. (5), which drops from 0.87 for $^{98}$Mo to 0.08 for $^{93}$Mo.

As we already mentioned, the less pronounced $S$ shape shows that the pairing correlations decrease when approaching the $N = 50$ shell gap. At the same time, the critical temperature for the quenching of pair correlations increases, which is the opposite of what one might expect. This effect can be explained by the increase in single-particle level spacing when approaching the $N = 50$ shell gap. We have already seen in the discussion in the previous section, that this increase, together with the weakening pairing correlations, which fail to push the nuclear ground state sufficiently down in energy, lead to a decrease in single particle entropy, see Figs. 7 and 12.

Therefore, the increase in critical temperature for the quenching of pairing correlations when approaching the $N = 50$ shell gap is due to the competition between the weakening pairing correlations and the increasing single-particle level spacing. Just as the weakened pairing correlations in odd nuclei cannot compensate for the effect of Pauli blocking on $T_c$, they cannot compensate for the effect of an increase in single-particle level spacing on $T_c$ when approaching a major shell gap.

## V. CONCLUSIONS

Levels in $^{93-98}$Mo in the excitation-energy region up to the neutron-separation energy were populated using ($^3$He,$\alpha\gamma$) and ($^3$He,$^3$He'$\gamma$) reactions. The level densities of $^{93-98}$Mo were determined from their corresponding primary $\gamma$-ray spectra. Within the canonical ensemble, thermodynamical observables were deduced from the level density; they display features consistent with signatures of a phase transition from a strongly pair-correlated phase to a phase without strong pairing correlations. This conclusion is supported by recent theoretical calculations within shell-model Monte Carlo simulations by Alhassid *et al.* [7, 8, 50], where it is shown that the expectation value of the pair operator decreases strongly around the critical temperature. However, we would like to point out, that other interpretations are not ex-cluded. Different mechanisms governing the thermodynamic properties of odd and even systems were studied. A simple, recently developed model for the investigation and classification of the pairing phase transition in hot nuclei has been employed and qualitative agreement with experimental data achieved. Using the saddle-point approximation the experimental level densities of even-even and odd-even systems are reproduced. Estimates for the critical temperature of the pairing-phase transition yield $T_c \sim 0.7$–$1.0$ MeV.

### Acknowledgments

Part of this work was performed under the auspices of the U.S. Department of Energy by the University of California, Lawrence Livermore National Laboratory under Contract W-7405-ENG-48. U.A., E.A., G.E.M., and A.V. acknowledge support from U.S. Department of Energy Grant No. DE-FG02-97-ER41042 and from the National Nuclear Security Administration under the Stockpile Stewardship Science Academic Alliances program through Department of Energy Research Grant Nos. DE-FG03-03-NA00074 and DE-FG03-03-NA00076. M.G., F.I., S.M., J.R., S.S., A.C.S., and S.Ø acknowledge financial support from the Norwegian Research Council (NFR).

[1] J. Bardeen, L. N. Cooper, and J. R. Schrieffer, Phys. Rev. **108**, 1175 (1957).
[2] M. Sano and S. Yamasaki, Prog. Theor. Phys. **29**, 397 (1963).
[3] A. L. Goodman, Nucl. Phys. **A352**, 45 (1981).
[4] L. G. Moretto, Nucl. Phys. **A185**, 145 (1972).
[5] K. Tanabe and K. Sugawara-Tanabe, Phys. Lett. **97B**, 337 (1980).
[6] S. Rombouts, K. Heyde, and N. Jachowicz, Phys. Rev. C **58**, 3295 (1998).
[7] S. Liu and Y. Alhassid, Phys. Rev. Lett. **87**, 022501 (2001).
[8] Y. Alhassid, G. F. Bertsch, and L. Fang, Phys. Rev. C **68**, 044322 (2003).
[9] J. L. Egido, L. M. Robledo, and V. Martin, Phys. Rev. Lett. **85**, 26 (2000).
[10] B. K. Agrawal, Tapas Sil, S. K. Samaddar, and J. N. De, Phys. Rev. C **63**, 024002 (2001).
[11] R. Rossignoli, N. Canosa, and P. Ring, Phys. Rev. Lett. **80**, 1853 (1998).
[12] A. Schiller, A. Bjerve, M. Guttormsen, M. Hjorth-Jensen, F. Ingebretsen, E. Melby, S. Messelt, J. Rekstad, S. Siem, and S. W.Ødegård, Phys. Rev. C **63**, 021306(R) (2001).
[13] E. Melby, L. Bergholt, M. Guttormsen, M. Hjorth-Jensen, F. Ingebretsen, S. Messelt, J. Rekstad, A. Schiller, S. Siem, and S.W. Ødegård, Phys. Rev. Lett. **83**, 3150 (1999).
[14] A. Schiller, L. Bergholt, M. Guttormsen, E. Melby, J. Rekstad, and S. Siem, Nucl. Instrum. Methods Phys. Res. A **447**, 498 (2000).
[15] G. A. Bartholomew, I. Bergqvist, E. D. Earle, and A. J. Ferguson, Can. J. Phys. **48**, 687 (1970).
[16] G. A. Bartholomew, E. D. Earle, A. J. Ferguson, J. W. Knowles, and M. A. Lone, Adv. Nucl. Phys. (N.Y.) **7**, 229 (1973).
[17] E. Melby, M. Guttormsen, J. Rekstad, A. Schiller, S. Siem, and A. Voinov, Phys. Rev. C **63**, 044309 (2001).
[18] A. Voinov, M. Guttormsen, E. Melby, J. Rekstad, A. Schiller, and S. Siem, Phys. Rev. C **63**, 044313 (2001).
[19] S. Siem, M. Guttormsen, K. Ingeberg, E. Melby, J. Rekstad, A. Schiller, and A. Voinov, Phys. Rev. C **65**, 044318 (2002).
[20] M. Guttormsen, E. Melby, J. Rekstad, S. Siem, A. Schiller, T. Lönnroth, and A. Voinov, J. Phys. G **29**, 263 (2003).
[21] A. Schiller *et al.*, Phys. Rev. C **68**, 054326 (2003).
[22] E. Tavukcu, Ph.D. thesis, North Carolina State University, 2002.
[23] M. Mayer, Elementary Theory of Nuclear Shell Structure (John Wiley & Sons, New York, 1955).
[24] S. Burger, and G. Heymann, Nucl. Phys. **A463**, 243 (1975).
[25] P. F. Mantica, A. E. Stuchbery, D. E. Groh, J. I. Prisciandaro, and M. P. Robinson, Phys. Rev. C **63**, 034312 (2001).
[26] M. Guttormsen, *et al.*, Phys. Rev. C **71**, 044307 (2005).
[27] M. Guttormsen, T. Ramsøy, and J. Rekstad, Nucl. Instrum. Methods Phys. Res. A **255**, 518 (1987).
[28] M. Guttormsen, T.S. Tveter, L. Bergholt, F. Ingebretsen, and J. Rekstad, Nucl. Instrum. Methods Phys. Res. A **374**, 371 (1996).
[29] L. Henden, L. Bergholt, M. Guttormsen, J. Rekstad, and T.S. Tveter, Nucl. Phys. **A589**, 249 (1995).
[30] D.M. Brink, Ph.D. thesis, Oxford University, 1955.
[31] P. Axel, Phys. Rev. **126**, 671 (1962).
[32] Data extracted using the NNDC On-Line Data Service from the ENSDF database, file revised as of Jan. 21, 2000.
[33] *Handbook for Calculations of Nuclear Reaction Data* (IAEA, Vienna, 1998).
[34] A. Gilbert, A.G.W. Cameron, Can. J. Phys. **43**, 1446 (1965).
[35] T. von Egidy, H.H. Schmidt, and A.N. Behkami, Nucl. Phys. **A481**, 189 (1988).
[36] G. Audi and A.H. Wapstra, Nucl. Phys. **A729**, 337 (2003).
[37] J. Dobaczewski, P. Magierski, W. Nazarewicz, W. Satuła, and Z. Szymański, Phys. Rev. C **63**, 024308 (2001).
[38] D. Bucurescu, T. von Egidy, J. Phys. G: Nucl. Part. Phys. **31**, S1675-S1680, (2005).
[39] R.B. Firestone, and V.S. Shirley, *Table of Isotopes*, 8th edition, Vol. I (New York: John Wiley & Sons, Inc., 1996).
[40] A.S. Iljinov, M.V. Mebel, N. Bianchi, E. De Sanc-

tis, C. Guaraldo, V. Lucherini, V. Muccifora, E. Polli, A.R. Reolon, and P. Rossi, Nucl. Phys. **A543**, 517 (1992), and references therein.

[41] M. Guttormsen, M. Hjorth-Jensen, E. Melby, J. Rekstad, A. Schiller, and S. Siem, Phys. Rev. C **63**, 044301 (2001).

[42] A. Schiller *et al.*, AIP Conf. Proc. **777**, 216 (2005).

[43] A. Schiller, M. Guttormsen, M. Hjorth-Jensen, J. Rekstad, and S. Siem, nucl-th/0306082.

[44] M. Guttormsen, A. Bagheri, R. Chankova, J. Rekstad, S. Siem, A. Schiller, and A. Voinov, Phys. Rev. C **68**, 064306 (2003).

[45] U. Agvaanluvsan *et al.*, Phys. Rev. C **70**, 054611 (2004).

[46] T.F. Wang, private communication.

[47] M. Guttormsen, M. Hjorth-Jensen, E. Melby, J. Rekstad, A. Schiller, and S. Siem, Phys. Rev. C **64**, 034319 (2002).

[48] A. Schiller, M. Guttormsen, M. Hjorth-Jensen, J. Rekstad, and S. Siem, Phys. Rev. C **66**, 024322 (2002).

[49] T. Døssing, *et al.*, Phys. Rev. Lett. **75**, 1276 (1995).

[50] H. Nakada and Y. Alhassid, Phys. Rev. Lett. **79**, 2939 (1997).

[51] K. Esashika, H. Nakada, and K. Tanabe, Phys. Rev. C **72**, 044303(2005).

TABLE I: Parameters used for the back-shifted Fermi-gas level density.

| Nucleus | $E_{\mathrm{pair}}$ (MeV) | $a$ (MeV$^{-1}$) | $C_1$ (MeV) | $B_n$ (MeV) | $D$ (eV) | $\rho(B_n)$ ($10^4$MeV$^{-1}$) | $\eta$ |
|---|---|---|---|---|---|---|---|
| $^{98}$Mo | 2.080 | 11.33 | -1.521 | 8.642 | 75 | 9.99 | 0.87 |
| $^{97}$Mo | 0.995 | 11.23 | -1.526 | 6.821 | 1050 | 3.10 | 0.65 |
| $^{96}$Mo | 2.138 | 11.13 | -1.531 | 9.154 | 105 | 7.18 | 0.46 |
| $^{95}$Mo | 1.047 | 11.03 | -1.537 | 7.367 | 1320 | 2.50 | 0.34 |
| $^{94}$Mo | 2.027 | 10.93 | -1.542 | 9.678 | | 6.20[a] | 0.25 |
| $^{93}$Mo | 0.899 | 10.83 | -1.547 | 8.067 | 2700 | 1.27 | 0.08 |

[a]Estimated from systematics, see text.

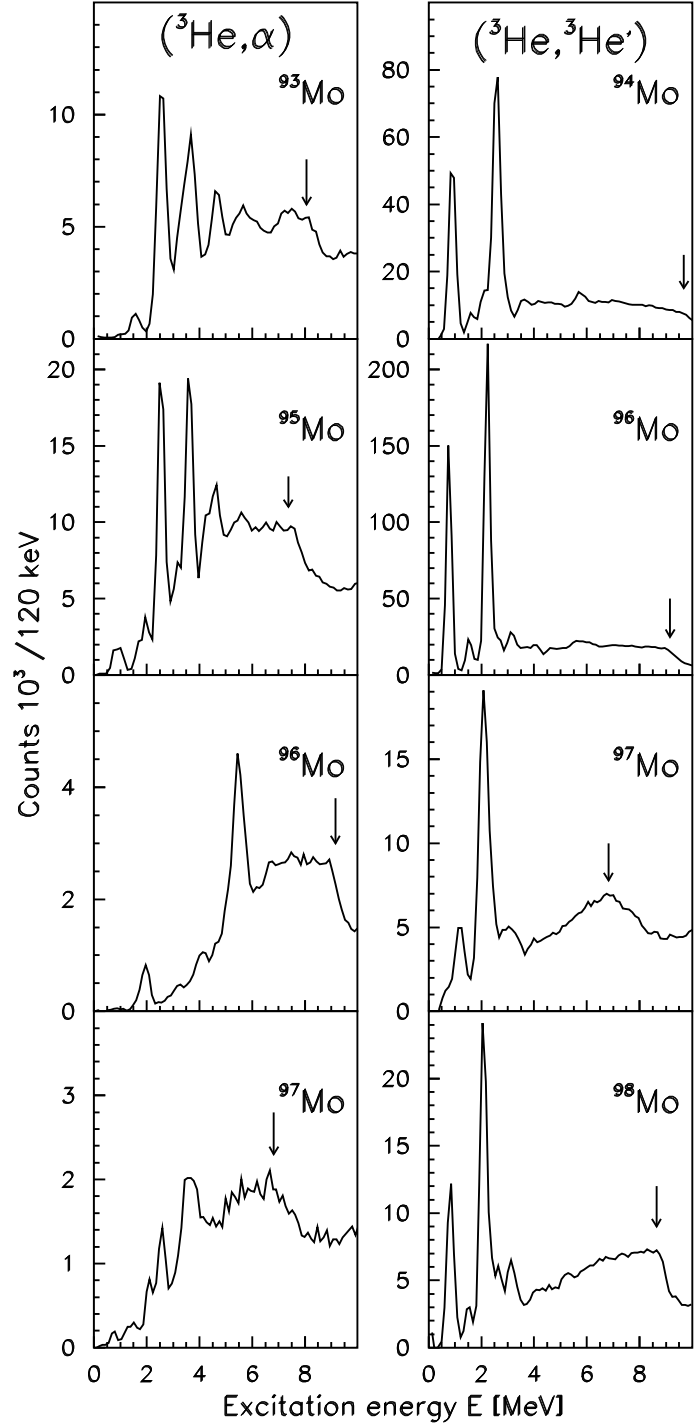

FIG. 1: Charged ejectile spectra for $^{93-98}$Mo in coincidence with $\gamma$-rays, labelled by the product nuclei. The arrows indicate the neutron separation energy $B_n$.

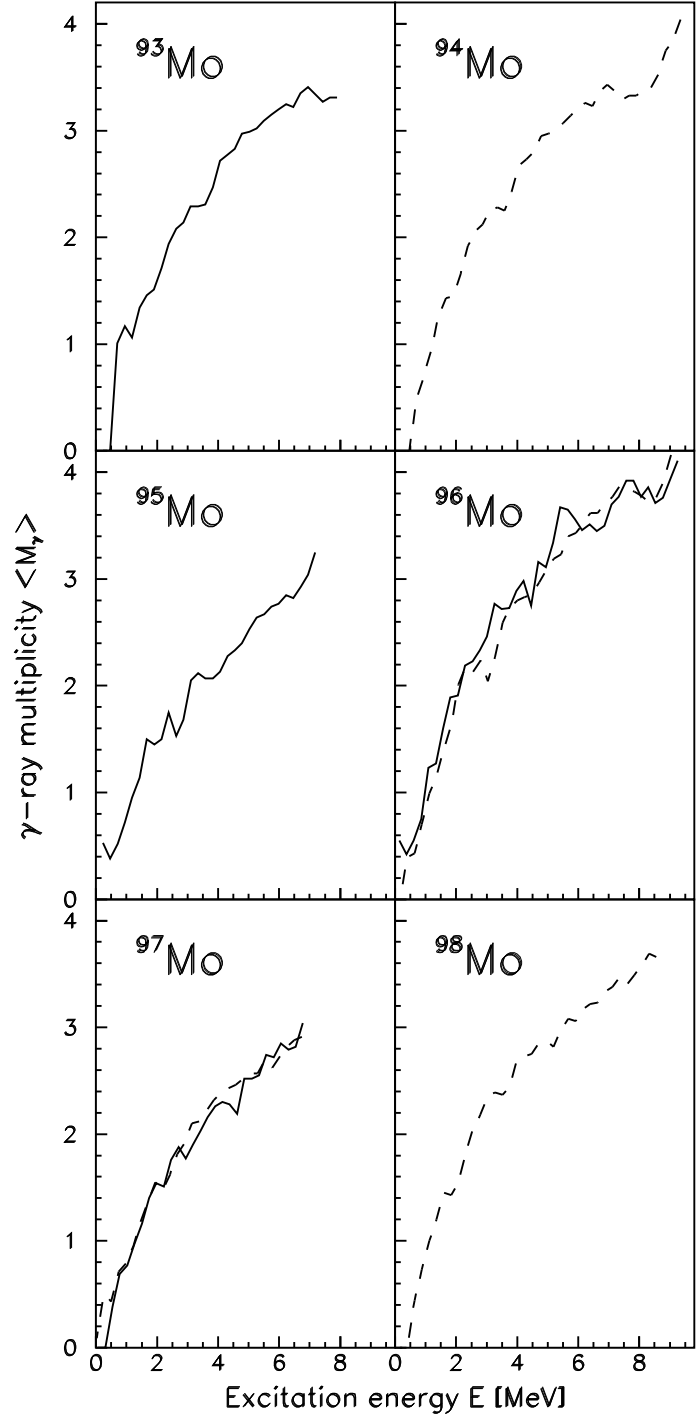

FIG. 2: γ-ray multiplicity evaluated by Eq. (1) versus excitation energy. The individual spectra are labelled by the product nuclei. Solid and dashed lines represent ($^3$He,α) and ($^3$He,$^3$He′) reactions, respectively.

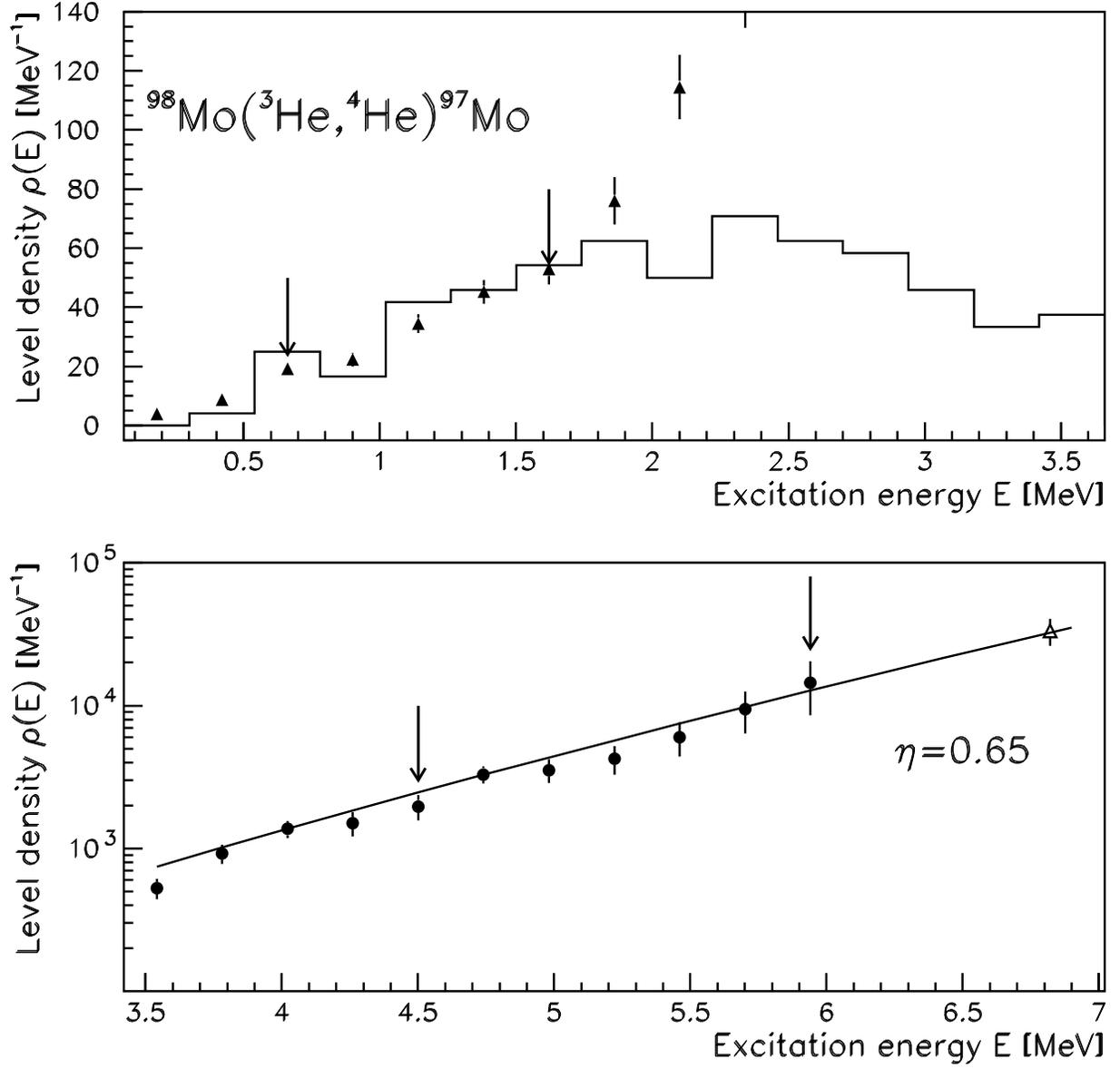

FIG. 3: Normalization procedure of the experimental level density (data points) of $^{97}$Mo. The data points between the arrows in the upper panel are normalized to known levels at low excitation energy (histograms). In the lower panel they are normalized to the level density at the neutron-separation energy (open triangle) using a Fermi-gas extrapolation (line).

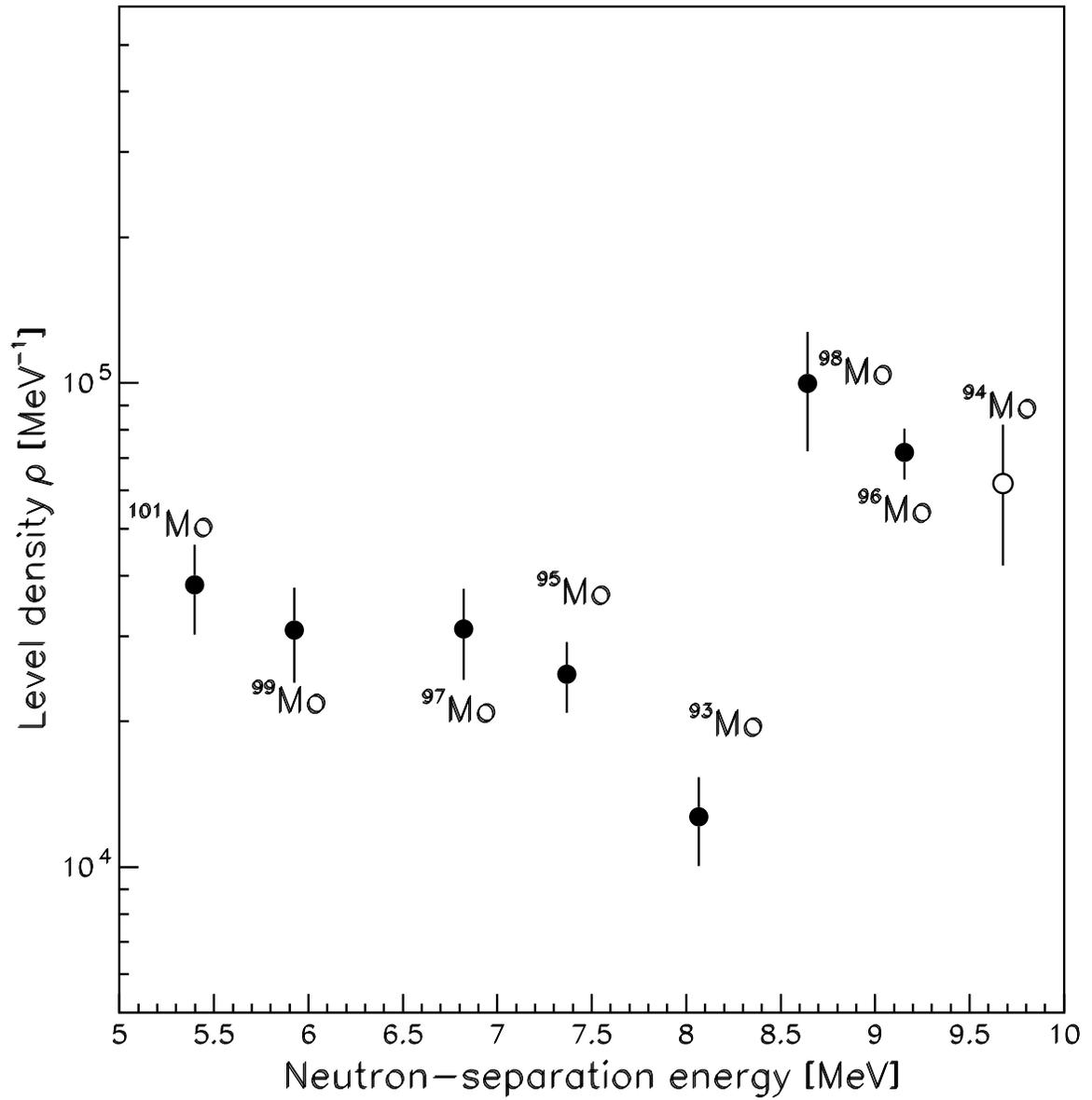

FIG. 4: Level densities at the neutron separation energy. The unknown level density of $^{94}$Mo (open circle) is estimated from the slope of the data points of the odd-mass molybdenum isotopes.

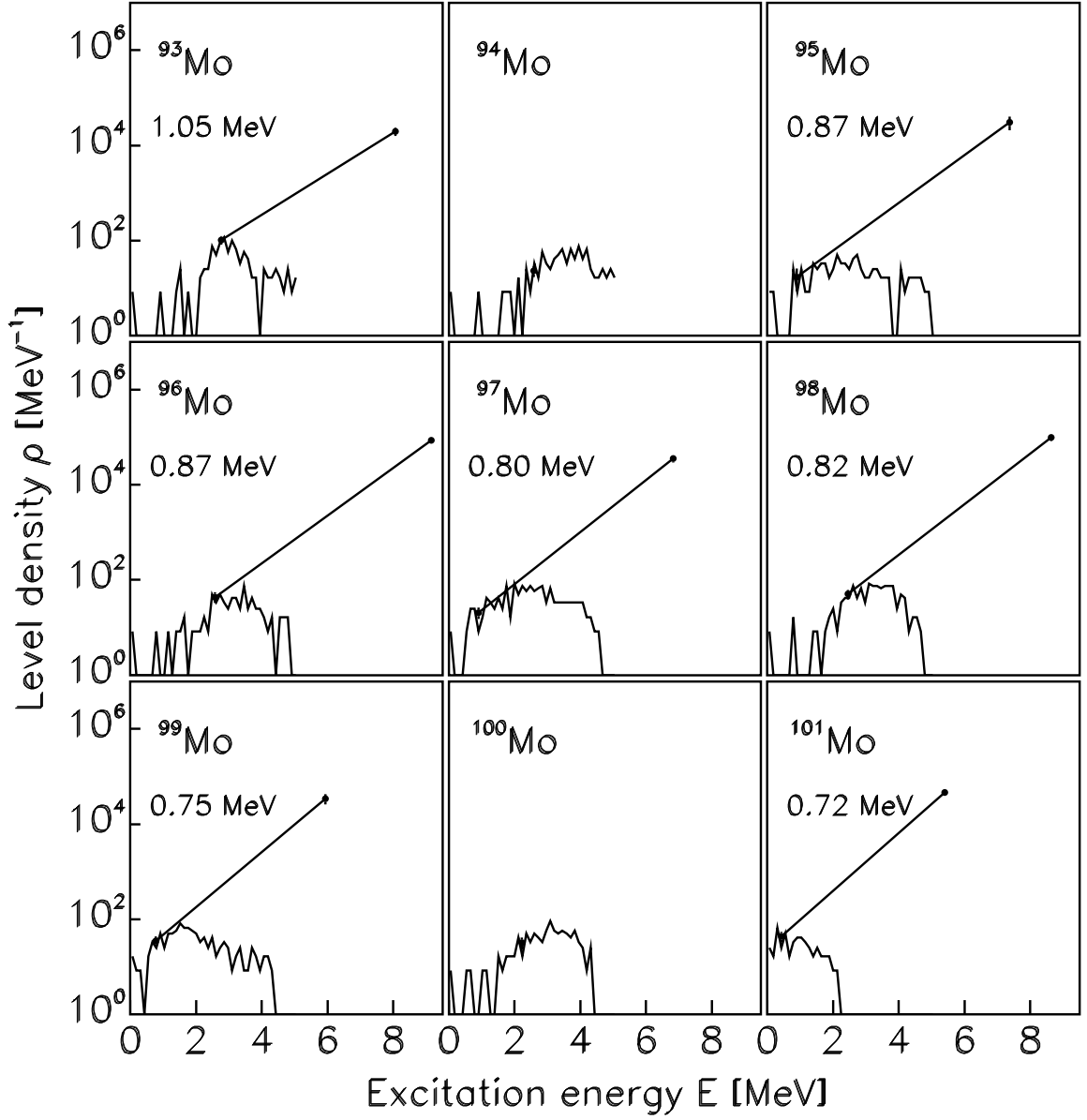

FIG. 5: Level density of nine molybdenum isotopes. The histograms represent levels from spectroscopy [39]. A straight line is drawn from these levels to the level density at the neutron-separation energy which is determined by average neutron resonance spacings. The line represents the constant-temperature level-density formula, see text.

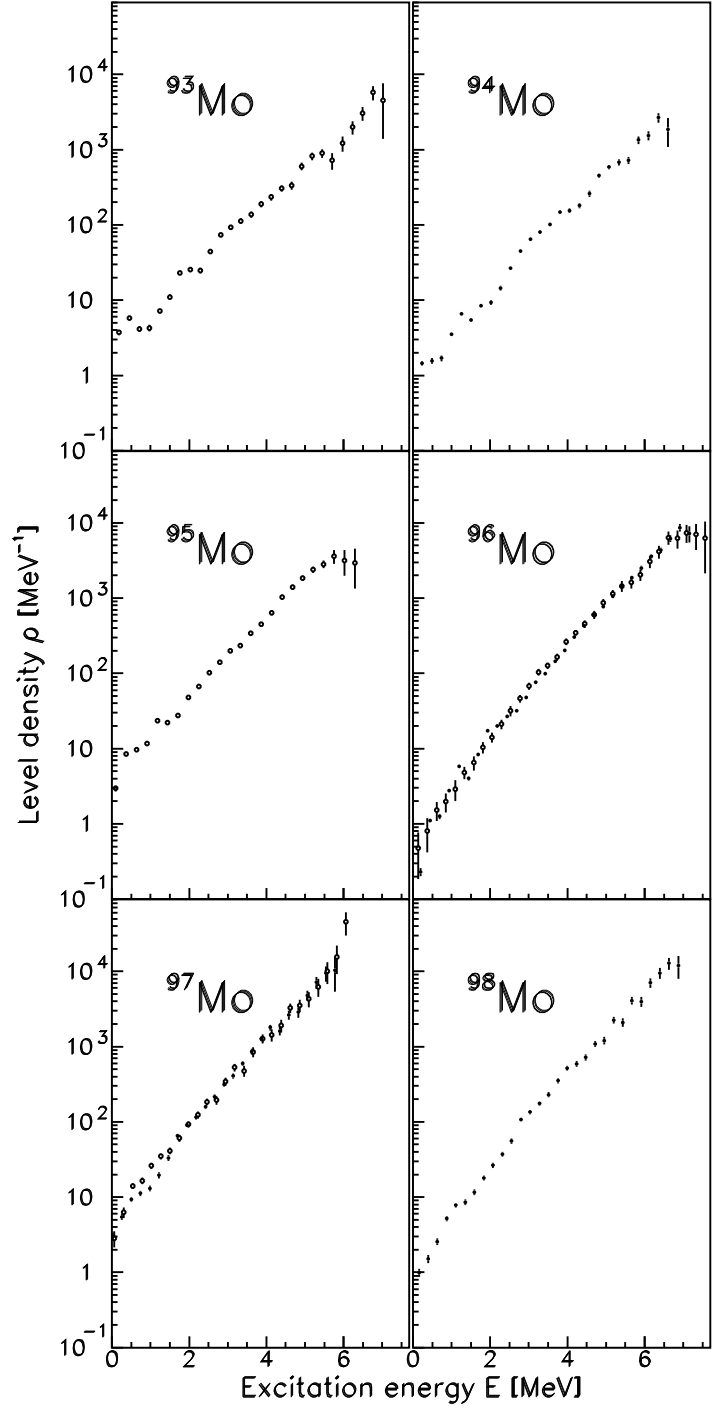

FIG. 6: Normalized level densities for $^{93-98}$Mo. The open and filled circles are data from the ($^3$He,$\alpha$) and ($^3$He,$^3$He') reactions, respectively.

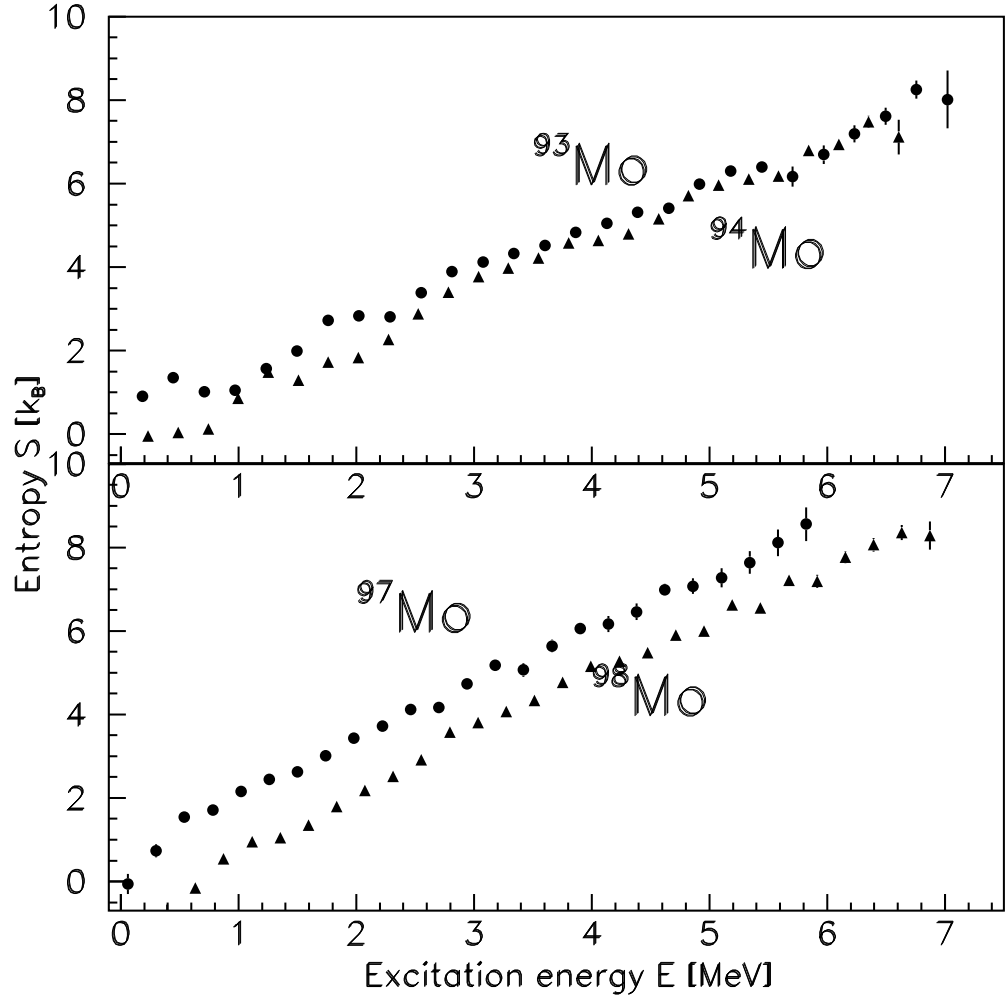

FIG. 7: Experimental entropy for $^{93,94}$Mo (upper panel) and $^{97,98}$Mo (lower panel) as function of excitation energy $E$.

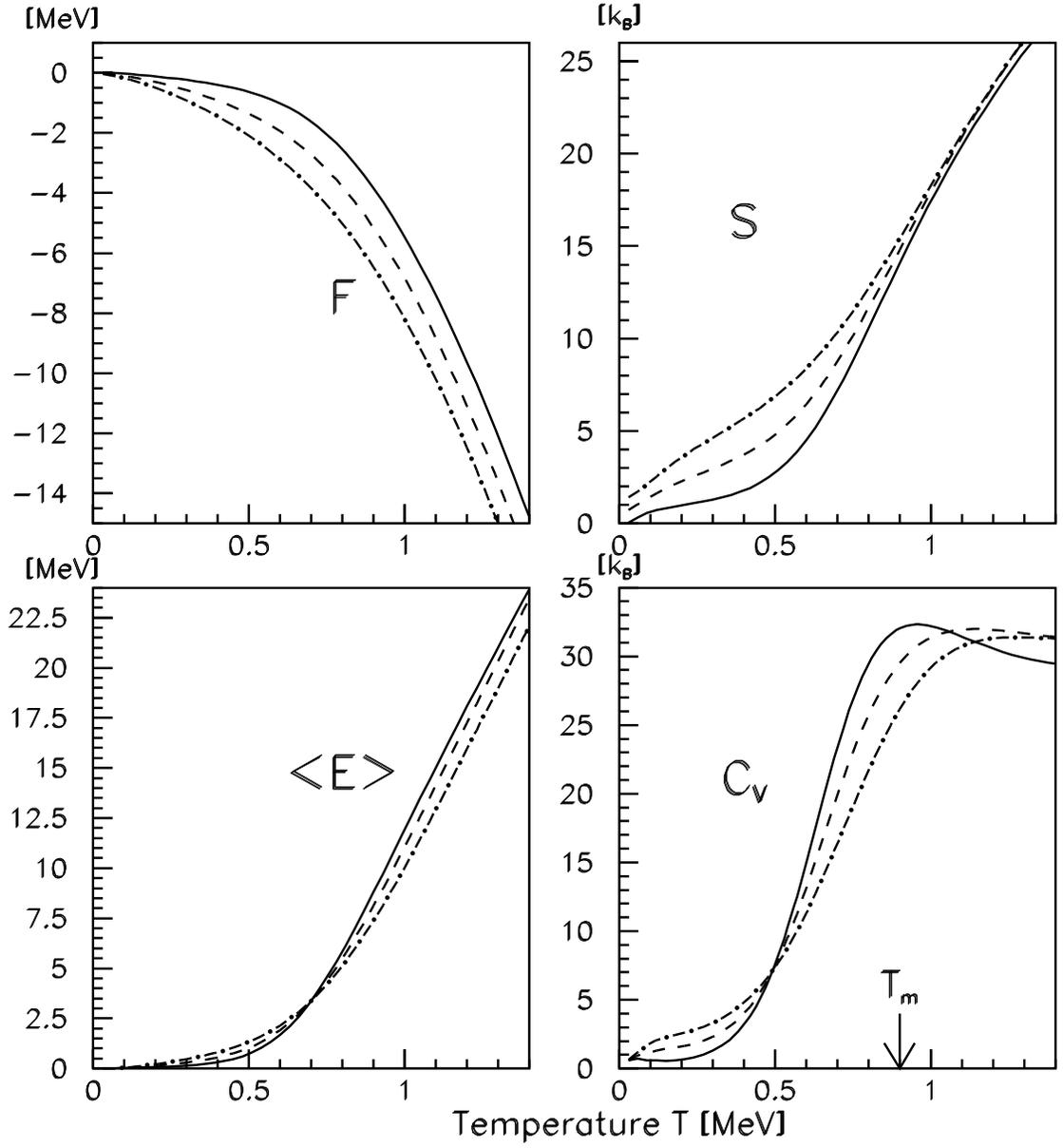

FIG. 8: Model calculation for nuclei around $^{96}$Mo. The four panels show the free energy $F$, the entropy $S$, the thermal excitation energy $\langle E \rangle$, and the heat capacity $C_V$ as a function of temperature $T$. The arrow at $T_m \sim 0.9$ MeV indicates the local maximum of $C_V$ where the pair-breaking process takes place in the even-even system. The same parameter set is used for even-even (solid lines), odd (dashed lines), and odd-odd systems (dash-dotted lines).

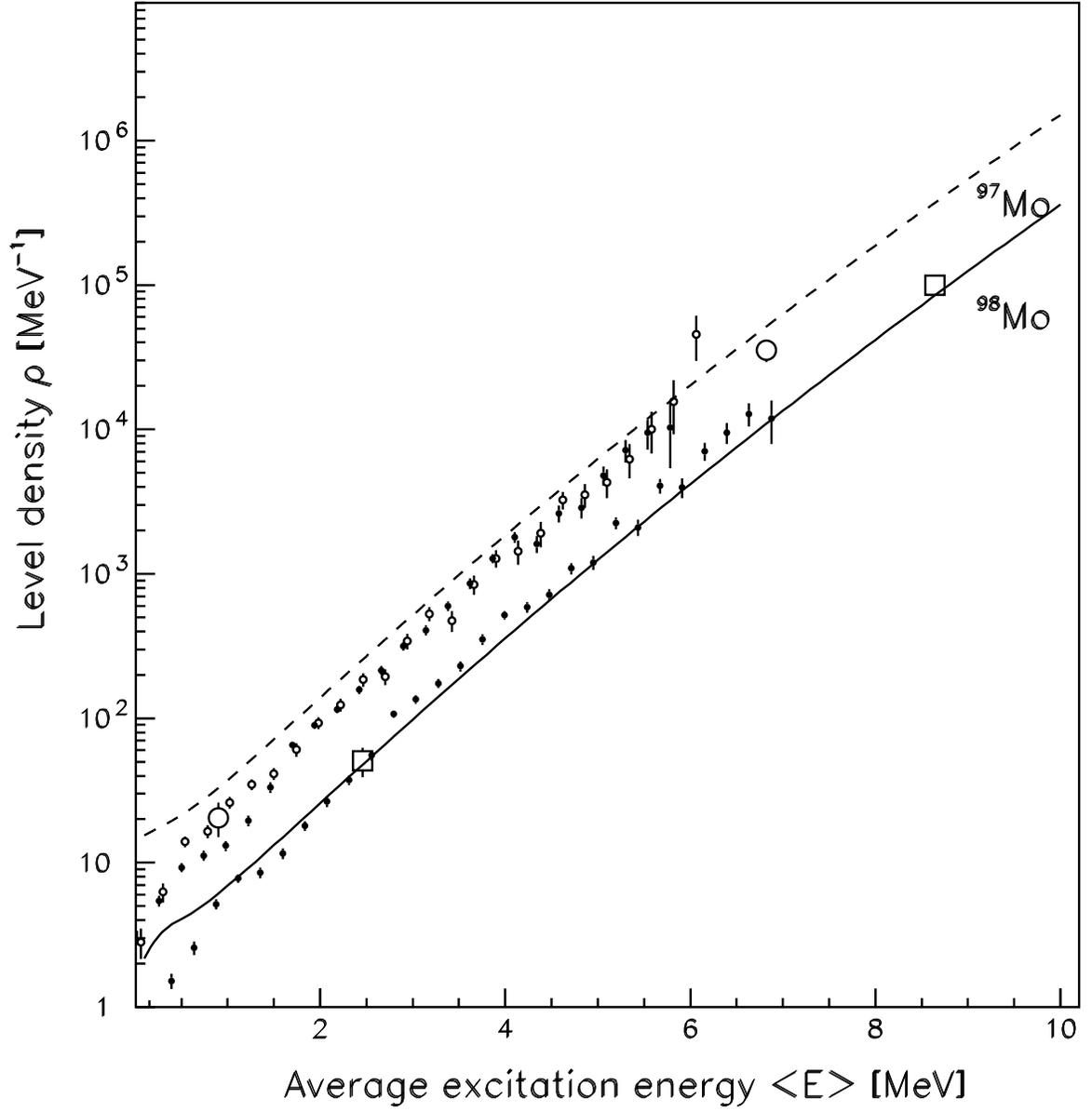

FIG. 9: Calculated level density of $^{98}$Mo (solid line) and $^{97}$Mo (dashed line) as function of average excitation energy $\langle E \rangle$. The big open circles and squares are experimental level-density anchor points from Ref. [41]. The small filled and open circles are experimental data points measured with the ($^3$He,$\alpha$) and ($^3$He,$^3$He$'$) reactions, respectively for the two isotopes.

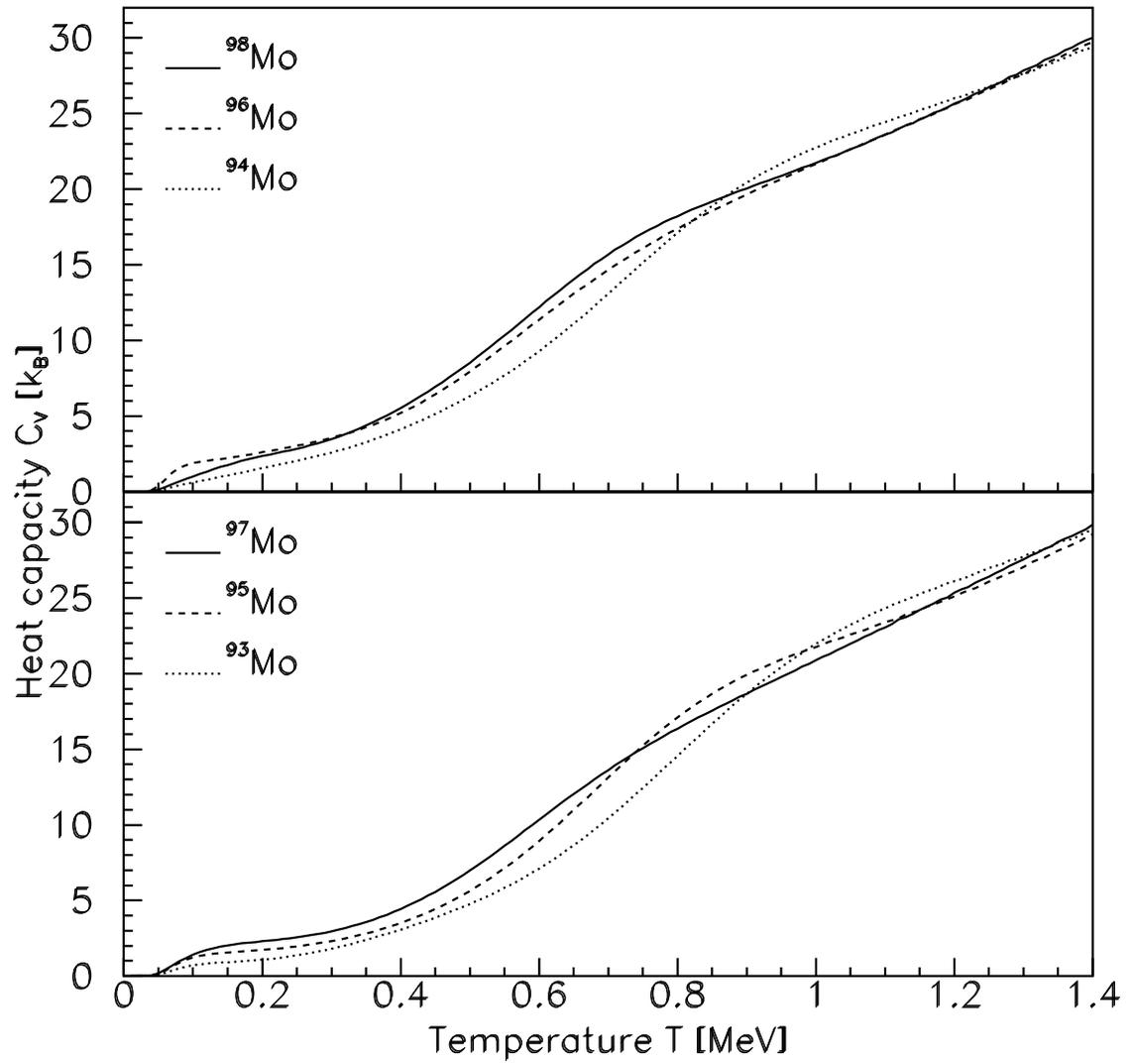

FIG. 10: Observed heat capacity as functions of temperature in the canonical ensemble for the even [94,96,98]Mo (upper panel) and odd [93,95,97]Mo (lower panel) nuclei.

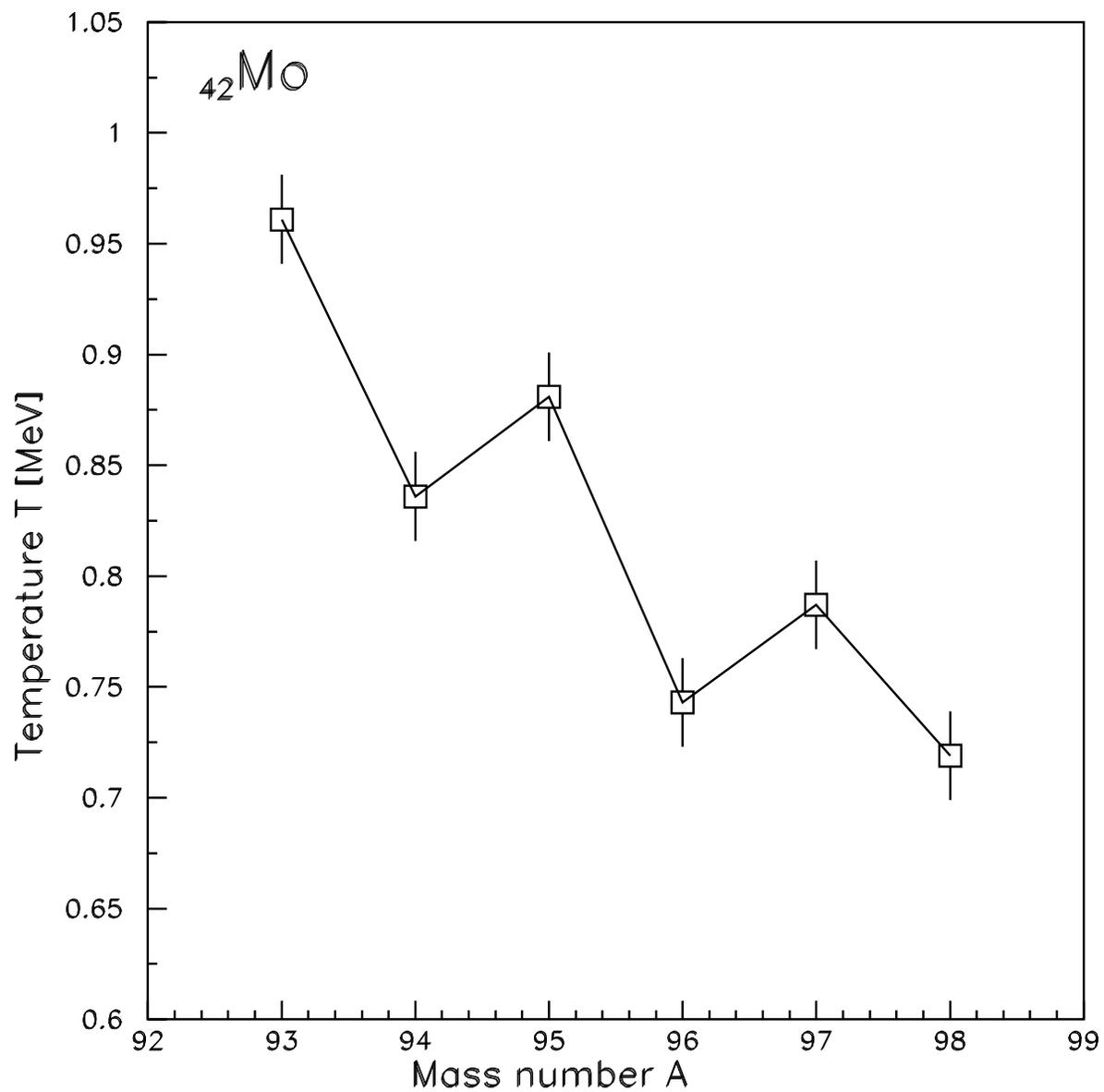

FIG. 11: Critical temperature for the quenching of pair correlations for $^{93-98}$Mo isotopes.

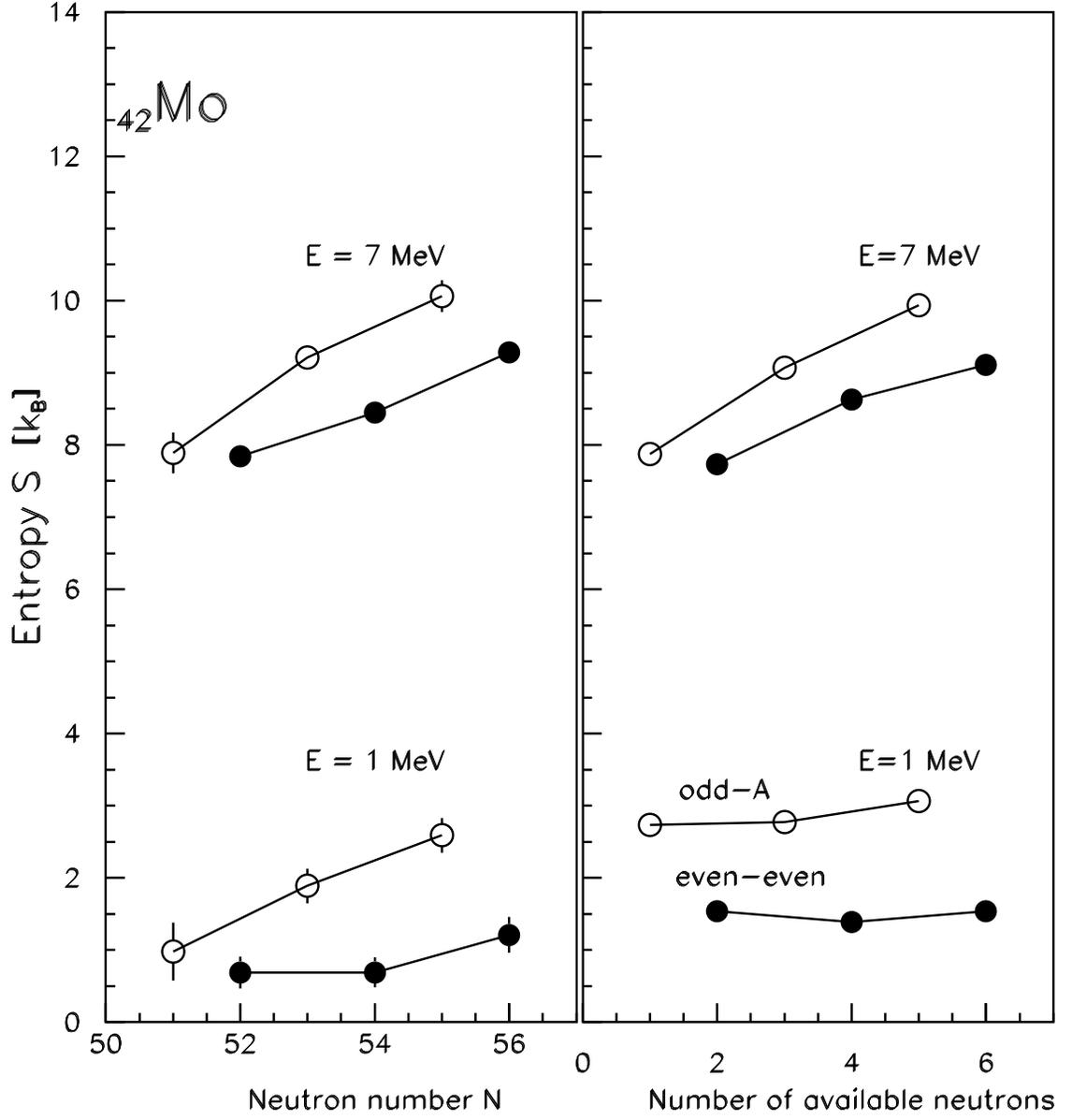

FIG. 12: Entropy extracted at 1 and 7 MeV excitation energy as function of neutron number $N$ (left panel) and number of available neutrons in the model (right panel) for odd-even (open circles) and even-even (filled circles) molybdenum isotopes.